\documentclass[aps,pra,floatfix,preprint,onecolumn,superscriptaddress]{revtex4-2}
\usepackage{graphicx}
\usepackage{amsmath, amsfonts, amssymb, bm}

\usepackage{amstext}
\usepackage{mathrsfs}
\usepackage{hyperref}
\usepackage{cleveref}

\usepackage{color}

\begin{document}
\title{Radiation by a laser-driven flying-focus electron wave packet}
\author{Antonino Di Piazza}
\email{a.dipiazza@rochester.edu}
\affiliation{Department of Physics and Astronomy, University of Rochester, Rochester, NY 14627, USA}
\affiliation{Laboratory for Laser Energetics, University of Rochester, Rochester, NY 14623, USA}
\author{Martin Formanek}
\affiliation{ELI Beamlines Facility, The Extreme Light Infrastructure ERIC, 252 41 Doln\'{i} B\v{r}e\v{z}any, Czech Republic}
\author{Dillon Ramsey}
\affiliation{Laboratory for Laser Energetics, University of Rochester, Rochester, NY 14623, USA}
\author{John P. Palastro}
\affiliation{Laboratory for Laser Energetics, University of Rochester, Rochester, NY 14623, USA}

\begin{abstract}
An exact solution of the Dirac equation in the presence of an arbitrary electromagnetic plane wave is found, which corresponds to a focused electron wave packet, with the focus of the wave packet moving at the speed of light in the opposite direction of the average momentum of the electron wave packet (unless the plane wave is so intense to reflect the electron). The photon spectrum emitted by such an electron wave packet in the presence of a linearly-polarized plane wave is studied both analytically and numerically. The spectrum is also compared with the one emitted by a single-momentum, plane-wave electron in the case of the electron being initially counter-propagating (on average for the flying-focus case) with the plane wave and within the locally-constant field approximation. It is found that if the electron flying-focus wave packet is focused beyond a Compton wavelength, the angular distribution of the emitted radiation along the magnetic field of the electromagnetic plane wave is broader than for an electron with definite momentum. Corresponding the maximum value of the photon yield on the transverse plane is smaller in the flying-focus electron case. This could represent an experimental signature of a laser-driven flying-focus electron wave packet.  
\end{abstract}

\maketitle
\thispagestyle{empty}

\section{Introduction}
The possibility of producing light waves with peculiar spacetime structures is attracting the attention of several groups as ``structured light'' can open new and unique frontiers of light-matter interaction (see, e.g., the two recent reviews \cite{Angelsky_2020,Wang_2021} and the references therein). Such investigations include, among others, the generation of light beams with non-zero orbital angular momentum, which are nowadays routinely produced in the laboratory (see Refs. \cite{Yao_2011,Zhang_2020_b} for comprehensive reviews in this field).

A notable example of structured light is represented by flying focus beams, i.e., beams whose focal velocity is independent of the group velocity and can be controlled externally \cite{Sainte-Marie_2017,Froula_2018}. Different techniques have been proposed to generate flying focus beams like the use of a chromatic lens combined with a chirped laser pulse (the chromatic flying focus) or an axiparabola combined with an echelon (the ultrashort flying focus) \cite{Palastro_2020,Ambat_2023}. The focal velocity can take arbitrary values independently of the laser group velocity and the focus can travel over distances much longer than a Rayleigh range. Flying focus beams offer unique possibilities to revisit established schemes and study regimes where they provide significant advantages as compared to, for example, fixed-focus Gaussian pulses. These include ionization waves of arbitrary velocity \cite{Turnbull_2018,Palastro_2018}, photon acceleration \cite{Howard_2019}, laser wakefield acceleration \cite{Palastro_2020}, vacuum electron acceleration \cite{Ramsey_2020}, transverse formation length enhancement in nonlinear Compton scattering \cite{Di_Piazza_2021}, nonlinear Thomson scattering \cite{Ramsey_2022}, radiation reaction \cite{Formanek_2022}, beam transport \cite{Formanek_2023} and vacuum-polarization effects \cite{Formanek_2024}. These same concepts have been applied to the formation of plasma waves that exhibit the same structure \cite{Palastro_2024_b}.

Structured ``electron'' pulses have been recently also investigated and ``electron vortices'', i.e., electron wave packets featuring non-zero orbital angular momentum \cite{Bu_2024}, have been produced in the laboratory \cite{Lloyd_2017}. In this context, we have recently shown that the Dirac equation in vacuum admits flying-focus like solutions, with arbitrary focal velocities \cite{Palastro_2024}. Indeed, the solutions of the Dirac equation can be related to the solutions of the solutions of the Klein-Gordon equation, whose structure is the same as the electromagnetic wave equation apart from a non-zero additional mass term.
 
In the present paper, we analyze the problem of a flying-focus electron in the case of the dressed Dirac equation in the presence of an arbitrary plane-wave electromagnetic field. After considering the case of an electron wave packet propagating in vacuum, we move to the more complicated case of an electron moving in the presence of an arbitrary plane-wave field. In this case, we find a new analytical solution of the dressed Dirac equation, which exactly describes a focused electron wave packet, whose focus corresponding to each momentum component of the wave packet moves at the speed of light in the opposite direction of the (subluminal) average velocity of the wave packet. In the vacuum case, i.e., in the case of a vanishing background plane wave, the solution reduces to the corresponding special solution in Ref. \cite{Palastro_2024} with the same focal velocity as here. Finally, we compute the probability of emission of a single photon by such an electron wave packet by taking into account exactly the presence of the background plane-wave field (nonlinear single Compton scattering) \cite{Di_Piazza_2012,Gonoskov_2022,Fedotov_2023}. The angular properties of the emitted radiation are studied both analytically and numerically and compared with the standard case of an incoming electron having a definite momentum.

\section{Notation}

For the sake of convenience, below we use units with $\hbar=c=\epsilon_0=1$ and the metric $\eta^{\mu\nu}=\text{diag}(+1,-1,-1,-1)$. In these units the fine-structure constant is $\alpha=e^2/4\pi\approx 1/137$.

As in the case of the flying focus light wave studied in Refs. \cite{Di_Piazza_2021,Formanek_2023}, it is convenient to use light-cone coordinates and, for the sake of definiteness, we assume that the focus moves along the positive $z$ direction. Thus, we define
\begin{align}
T=\frac{t+z}{2}, && \bm{x}_{\perp}=(x,y), && \phi=t-z
\end{align}
for a spacetime point with coordinates $x^{\mu}=(t,\bm{x})=(t,x,y,z)$. The light-cone components of an arbitrary four-vector $v^{\mu}=(v_0,\bm{v})$ are defined as $v_+=(v_0+v_z)/2$, $\bm{v}_{\perp}=(v_x,v_y)$, and $v_-=v_0-v_z$. The same definition is adopted for the Dirac gamma matrices $\gamma^{\mu}=(\gamma^0,\bm{\gamma})$: $\gamma_+=(\gamma^0+\gamma^3)/2$, $\bm{\gamma}_{\perp}=(\gamma^1,\gamma^2)$, and $\gamma_-=\gamma^0-\gamma^3$, which are chosen in the standard representation. The derivatives with respect to the light-cone coordinates $T$ and $\phi$ are
\begin{align}
\frac{\partial}{\partial T}=\frac{\partial}{\partial t}+\frac{\partial}{\partial z}, && \frac{\partial}{\partial \phi}=\frac{1}{2}\left(\frac{\partial}{\partial t}-\frac{\partial}{\partial z}\right),
\end{align}
whereas the derivatives with respect to the transverse coordinates form the two-dimensional vector $\bm{\nabla}_{\perp}=\partial/\partial \bm{x}_{\perp}=(\partial/\partial x, \partial/\partial y)$. The scalar product between two four-vectors $u^{\mu}$ and $v^{\mu}$ is expressed in light-cone components as
\begin{equation}
(uv)=u_+v_-+u_-v_+-\bm{u}_{\perp}\cdot\bm{v}_{\perp}
\end{equation}
and the four-divergence of a vector field $F^{\mu}(x)$ as
\begin{equation}
\label{Four_div}
\partial_{\mu} F^{\mu}=\frac{\partial F_+}{\partial T}+\frac{\partial F_-}{\partial \phi}+\bm{\nabla}_{\perp}\cdot\bm{F}_{\perp}.
\end{equation}

In agreement with the above definitions, it is convenient to introduce the four-dimensional quantities: $n^{\mu}=(1,\bm{n})$ and $\tilde{n}^{\mu}=(1,-\bm{n})/2$, with $\bm{n}=(0,0,1)$ being the unit vector along the $z$ direction, and $a_j^{\mu}=(0,\bm{a}_j)$, where $j=1,2$, with $\bm{a}_1$ and $\bm{a}_2$ being the unit vectors along the $x$ and the $y$ direction, respectively. The four-dimensional quantities $n^{\mu}$, $\tilde{n}^{\mu}$, and $a^{\mu}_j$ fulfill the completeness relation: $\eta^{\mu\nu}=n^{\mu}\tilde{n}^{\nu}+\tilde{n}^{\mu}n^{\nu}-a_1^{\mu}a_1^{\nu}-a_2^{\mu}a_2^{\nu}$ (note that $(n\tilde{n})=1$ and $(a_1a_1)=(a_2a_2)=-1$, whereas all other possible scalar products among $n^{\mu}$, $\tilde{n}^{\mu}$, and $a^{\mu}_j$ vanish). By using the quantities $n^{\mu}$, $\tilde{n}^{\mu}$, and $a^{\mu}_j$ one can express the light-cone coordinates as $T=(\tilde{n}x)$, $\bm{x}_{\perp}=-((xa_1),(xa_2))$, and $\phi=(nx)$. Analogously, the light-cone components of an arbitrary four-vector $v^{\mu}=(v^0,\bm{v})$ are given by $v_+=(\tilde{n}v)$, $\bm{v}_{\perp}=-((va_1),(va_2))$, and $v_-=(nv)$, whereas, by using the ``hat'' notation for the contraction of a four-vector and the four Dirac gamma matrices, one obtains $\gamma_+=\hat{\tilde{n}}$, $\bm{\gamma}_{\perp}=-(\hat{a}_1,\hat{a}_2)$, and $\gamma_-=\hat{n}$. Concerning the derivatives with respect to the coordinates, the following relations hold: $\partial/\partial T=\partial_T=(n\partial)$, $\bm{\nabla}_{\perp}=((a_1\partial),(a_2\partial))$, and $\partial/\partial\phi=\partial_{\phi}=(\tilde{n}\partial)$. 

Concerning the background electromagnetic field, we assume that it propagates along the positive $z$ direction and that is described by the four-vector potential $A^{\mu}(\phi)$, which vanishes at $\phi\to\pm\infty$ and which satisfies the Lorenz-gauge condition $\partial_{\mu}A^{\mu}(\phi)=(nA'(\phi))=0$, such that $(nA(\phi))=0$. In the plane-wave case it is always possible within the Lorenz gauge to choose the four-vector potential as $A^{\mu}(\phi)=(0,\bm{A}_{\perp}(\phi))$, with $\bm{A}_{\perp}(\phi)=A_0[\psi_1(\phi)\bm{a}_1+\psi_2(\phi)\bm{a}_2]$. Here, $A_0$ is a constant which we will relate below to the amplitude of the plane wave and $\psi_j(\phi)$ denotes the pulse shape function along the $j$th direction, with $j=1,2$.

We recall that in the presence of a plane-wave electromagnetic field propagating along the positive $z$ direction, the light-cone minus and perpendicular momentum components are conserved, and they are convenient quantum numbers to construct the electron wave packets in configuration space. 

\section{Flying focus solution of the Dirac equation in an intense plane wave}
In this section, we derive an exact solution of the Dirac equation in the presence of the plane wave $A^{\mu}(\phi)$, which corresponds to a focused electron wave packet whose central momentum points along the negative $z$ direction, whereas the focus of each component of the wave packet moves along the positive $z$ direction at the speed of light.

For the sake of clarity, we will start from the solution of the Klein-Gordon equation in vacuum, then we proceed to the case of the Dirac equation in vacuum, and we conclude with the most complicated case of the Dirac equation in the presence of the plane wave $A^{\mu}(\phi)$.

By indicating as $m$ the electron mass, the Klein-Gordon equation in vacuum (for a hypothetical scalar particle with the electron mass)
\begin{equation}
(\partial^2+m^2)\varphi=0,
\end{equation}
with $\partial^2=\partial_{\mu}\partial^{\mu}$, can be written in light-cone coordinates as
\begin{equation}
(2\partial_T\partial_{\phi}-\bm{\nabla}_{\perp}^2+m^2)\varphi=0.
\end{equation}
Here, $\varphi(\phi,T,\bm{x}_{\perp})$ is the function, describing the ``spinless'' electron wave packet. Analogously as in Ref. \cite{Di_Piazza_2021}, we pass to the Fourier transform
\begin{equation}
\varphi(\phi,T,\bm{x}_{\perp})=\int\frac{dp_-}{2\pi}e^{-ip_-T}\tilde{\varphi}(\phi,p_-,\bm{x}_{\perp}),
\end{equation}
where the integral is, for the moment, assumed to extend from $-\infty$ to $+\infty$ as we ignore the physical meaning of the quantity $p_-$, which is merely an integration variable here.

The function  $\tilde{\varphi}(\phi,p_-,\bm{x}_{\perp})$ satisfies the paraxial-like equation
\begin{equation}
(-2ip_-\partial_{\phi}-\bm{\nabla}_{\perp}^2+m^2)\tilde{\varphi}=0.
\end{equation}
In order to reduce this equation to the electromagnetic, massless case \cite{Di_Piazza_2021}, we set
\begin{equation}
\tilde{\varphi}(\phi,p_-,\bm{x}_{\perp})=e^{-im^2\phi/2p_-}\Phi(\phi,p_-,\bm{x}_{\perp}),
\end{equation}
such that function $\Phi(\phi,p_-,\bm{x}_{\perp})$ satisfies the equation:
\begin{equation}
(-2ip_-\partial_{\phi}-\bm{\nabla}_{\perp}^2)\Phi=0.
\end{equation}
At this point, we can use the result in Ref. \cite{Di_Piazza_2021} and write
\begin{equation}
\Phi(\phi,p_-,\bm{x}_{\perp})=f(p_-)\frac{\sigma^2}{\Sigma^2(\phi)}e^{-\frac{\bm{x}_{\perp}^2}{2\Sigma^2(\phi)}},
\end{equation}
where $f(p_-)$ is an arbitrary function, 
\begin{equation}
\label{Sigma_sq}
\Sigma^2(\phi)=\sigma^2\left(1+i\frac{\phi}{l_{p_-}}\right)=\sigma^2+i\frac{\phi}{p_-},
\end{equation}
with $\sigma$ being the waist radius of the wave packet, and $l_{p_-}=p_-\sigma^2$ representing the Rayleigh length of the corresponding wave-packet component. The expression of $\Phi(\phi,p_-,\bm{x}_{\perp})$ indicates that the position of the focus, i.e., the minimal value of the function $|\Sigma(\phi)|^2$, is at $\phi=0$, i.e., it moves at the speed of light along the positive $z$ direction. It is worth noting any Laguerre-Gaussian or Hermite Gaussian mode (or more generally any solution to the paraxial wave equation) could be considered here, as well (see also Ref. \cite{Formanek_2023}). 

In summary, the exact solution of the Klein-Gordon equation, corresponding to a wave packet with a focus of radius $\sigma$ and moving at the speed of light along the positive $z$ direction is given by
\begin{equation}
\label{KG_Sol_Conf}
\varphi_{FF}(\phi,T,\bm{x}_{\perp})=\int\frac{dp_-}{2\pi}f(p_-)e^{-i\left(\frac{m^2}{2p_-}\phi+p_-T\right)}\frac{\sigma^2}{\Sigma^2(\phi)}e^{-\frac{\bm{x}_{\perp}^2}{2\Sigma^2(\phi)}}.
\end{equation}
It is convenient for what follows to express the function of $\bm{x}_{\perp}$ in the integrand via a Fourier integral:
\begin{equation}
\frac{\sigma^2}{\Sigma^2(\phi)}e^{-\frac{\bm{x}_{\perp}^2}{2\Sigma^2(\phi)}}=2\pi\sigma^2\int\frac{d^2\bm{p}_{\perp}}{(2\pi)^2}e^{i\bm{p}_{\perp}\cdot\bm{x}_{\perp}}e^{-\bm{p}_{\perp}^2\Sigma^2(\phi)/2}.
\end{equation}
In this way, the solution $\varphi_{FF}(\phi,T,\bm{x}_{\perp})$ can be simply written as
\begin{equation}
\label{KG_Sol}
\varphi_{FF}(\phi,T,\bm{x}_{\perp})=2\pi\sigma^2\int\frac{dp_-}{2\pi}\int\frac{d^2\bm{p}_{\perp}}{(2\pi)^2}f(p_-)e^{-i(px)}e^{-\bm{p}_{\perp}^2\sigma^2/2},
\end{equation}
where $p^{\mu}$ is the on-shell four-momentum with light-cone components $p_-$, $\bm{p}_{\perp}$, and $p_+=(m^2+\bm{p}^2_{\perp})/2p_-$, which also provides the physical meaning of the integration variable $p_-$. Here, it is understood that the function $f(p_-)$ is chosen in such a way that only positive values of $p_-$ contribute to the wave packet \footnote{The condition $p_->0$ guarantees that the electron energies under consideration are positive. If we wanted to consider wave packets only including positive values of the momentum $p_z$, then the more restrictive condition $p_->\sqrt{m^2+\bm{p}^2_{\perp}}$ should be fulfilled. We will see blow that in the regime of interest here the latter condition is well satisfied.}. Notice that in the limit $\sigma\to\infty$, the function $\sigma^2\exp(-\bm{p}_{\perp}^2\sigma^2/2)/2\pi$ tends to the delta function $\delta^2(\bm{p}_{\perp})$, such that the resulting wave packet has only components propagating exactly along the $z$ direction. The solution in Eq. (\ref{KG_Sol}) also indicates how to construct the wave packet by appropriately combining plane waves with a Gaussian distribution of the transverse momenta.

The above flying-focus solution in Eq. (\ref{KG_Sol}) gives already a hint on how to write the solution of the Dirac equation
\begin{equation}
(i\gamma^{\mu}\partial_{\mu}-m)\psi=0
\end{equation}
in vacuum, where $\psi(\phi,T,\bm{x}_{\perp})$ is a four-dimensional spinor. In fact, a class of solutions $\psi_{FF,s}(\phi,T,\bm{x}_{\perp})$ can be written as
\begin{equation}
\label{Dirac_Sol}
\begin{split}
\psi_{FF,s}(\phi,T,\bm{x}_{\perp})&=2\pi\sigma^2\frac{i\gamma^{\mu}\partial_{\mu}+m}{2m}\int\frac{dp_-}{2\pi}\int\frac{d^2\bm{p}_{\perp}}{(2\pi)^2}f(p_-)e^{-i(px)}e^{-\bm{p}_{\perp}^2\sigma^2/2}u_s(\bm{p})\\
&=2\pi\sigma^2\int\frac{dp_-}{2\pi}\int\frac{d^2\bm{p}_{\perp}}{(2\pi)^2}f(p_-)e^{-i(px)}e^{-\bm{p}_{\perp}^2\sigma^2/2}\frac{\hat{p}+m}{2m}u_s(\bm{p})\\
&=2\pi\sigma^2\int\frac{dp_-}{2\pi}\int\frac{d^2\bm{p}_{\perp}}{(2\pi)^2}f(p_-)e^{-i(px)}e^{-\bm{p}_{\perp}^2\sigma^2/2}u_s(\bm{p}),
\end{split}
\end{equation}
where $u_s(\bm{p})$ is the constant, positive-energy spinor solution of the equation $\hat{p}u_s(\bm{p})=mu_s(\bm{p})$ with spin quantum number $s$ referring to the spin orientation in the rest frame of the electron and normalized as $\bar{u}_s(\bm{p})u_s(\bm{p})=2m$ (the Dirac conjugated $\bar{\psi}$ of an arbitrary four-dimensional spinor $\psi$ is defined as $\bar{\psi}=\psi^{\dag}\gamma^0$) \cite{Landau_b_4_1982}. Note that, although the state $\psi_{FF,s}(\phi,T,\bm{x}_{\perp})$ is indicated to feature the quantum number $s$, it is not an eigenstate of any rest-frame spin operator, as it is clear from the fact that it is a linear combination of states with different momenta, such that a single rest frame cannot be defined.

An important observation is in order. Due to the presence of the projector $\hat{p}+m$, the spinor $\psi_{FF,s}(\phi,T,\bm{x}_{\perp})$ would be a solution of the Dirac equation for an arbitrary constant spinor instead of $u_s(\bm{p})$. In this respect, an alternative possibility would have been to use the spinor $u_s(\bm{p}_{\parallel})$, where $\bm{p}_{\parallel}=(0,0,p_z)=(0,0,m^2/2p_--p_-/2)$, with corresponding energy $\varepsilon_{p_{\parallel}}=m^2/2p_-+p_-/2$. The spinor $u_s(\bm{p}_{\parallel})$ is independent of $\bm{p}_{\perp}$ and this, in analogy with the scalar case, guarantees that the structure of the wave packet resembles that of a flying focus electron with a Gaussian transverse profile, once the integral in $\bm{p}_{\perp}$ is taken. The disadvantage, however, of using $u_s(\bm{p}_{\parallel})$ is that its structure would change under a general Lorentz boost. In other words, we would have implicitly made a choice of the reference system where the structure of $\psi_{FF,s}(\phi,T,\bm{x}_{\perp})$ would have been that corresponding to the choice of $u_s(\bm{p}_{\parallel})$. As a consequence, the obtained results using that $\psi_{FF,s}(\phi,T,\bm{x}_{\perp})$ would have not been manifestly Lorentz covariant. In this respect, the manifest Lorentz covariance is instead guaranteed by the choice of $u_s(\bm{p})$, whose structure is independent of the reference system by construction.

The problem now arises of whether the spinor $\psi_{FF,s}(\phi,T,\bm{x}_{\perp})$ in Eq. (\ref{Dirac_Sol}) has a flying-focus structure with a Gaussian transverse profile like analogous to the solution of the Klein-Gordon equation. By using the identity
\begin{equation}
u_s(\bm{p})=\frac{\hat{p}+m}{\sqrt{2m(\varepsilon_p+m)}}u_s(\bm{0}),
\end{equation}
with $u_s(\bm{0})$ being independent of $\bm{p}_{\perp}$, it is easily seen that the presence of the energy $\varepsilon_p=\sqrt{m^2+\bm{p}^2}=p_++p_-/2=(m^2+\bm{p}_{\perp}^2)/2p_-+p_-/2$ in the denominator prevents one from taking the integral in $\bm{p}_{\perp}$ analytically. However, it can be shown that in the important case (see also below) of an ultrarelativistic electron with typical values of $p_-$ much larger than $m$ and for values of the transverse size of the wave packet $\sigma$ such that $\sigma\lesssim \lambda_C=1/m$, the integral in $\bm{p}_{\perp}$ can be approximately taken analytically and the resulting structure of the wave packet is again of a flying-focus electron with transverse Gaussian profile. This is essentially achieved by using the identity
\begin{equation}
\frac{1}{\sqrt{\varepsilon_p+m}}=\int \frac{d\eta}{\sqrt{\pi}}e^{-(\varepsilon_p+m)\eta^2}=\int \frac{d\eta}{\sqrt{\pi}}e^{-\left(\frac{m^2+\bm{p}_{\perp}^2}{2p_-}+\frac{p_-}{2}+m\right)\eta^2},
\end{equation}
which makes the resulting integral in $\bm{p}_{\perp}$ Gaussian:
\begin{equation}
\begin{split}
\psi_{FF,s}(\phi,T,\bm{x}_{\perp})&=2\pi\sigma^2\int\frac{dp_-}{2\pi}\int\frac{d^2\bm{p}_{\perp}}{(2\pi)^2}\int d\eta f(p_-)e^{-i\left(\frac{m^2}{2p_-}\phi+p_-T\right)}e^{-\left(\frac{\sigma^2}{2}+\frac{\eta^2}{2p_-}+i\frac{\phi}{2p_-}\right)\bm{p}_{\perp}^2+i\bm{p}_{\perp}\cdot\bm{x}_{\perp}}\\
&\quad\times e^{-\left(\frac{m^2}{2p_-}+\frac{p_-}{2}+m\right)\eta^2}\frac{\hat{p}+m}{\sqrt{2\pi m}}u_s(\bm{0}).
\end{split}
\end{equation}
This equation is still exact and the integrand already shows that the values of $\eta$ mostly contributing to the integral are such that $\eta\lesssim \sqrt{2p_-}/(p_-+m)$. Looking at the coefficient of $\bm{p}_{\perp}^2$ in the exponent, we conclude that the quantity $\eta^2/2p_-$ is negligible as compared to the quantity $\sigma^2/2$ if $\sigma^2\gg 2/(p_-+m)^2$, which is the case under the conditions mentioned above.

By neglecting the quantity $\eta^2/2p_-$ as compared to the quantity $\sigma^2/2$ and after taking the Gaussian integral in $\bm{p}_{\perp}$, also the integral in $\eta$ is Gaussian and can be taken analytically. The final result is
\begin{equation}
\begin{split}
\psi_{FF,s}(\phi,T,\bm{x}_{\perp})&\approx\int\frac{dp_-}{2\pi}f(p_-)e^{-i\left(\frac{m^2}{2p_-}\phi+p_-T\right)}\frac{\sigma^2}{\Sigma^2(\phi)}e^{-\frac{\bm{x}_{\perp}^2}{2\Sigma^2(\phi)}}\\
&\quad\times\frac{\sqrt{p_-/m}}{p_-+m}\left[\frac{m^2\Sigma^4(\phi)+2\Sigma^2(\phi)-\bm{x}_{\perp}^2}{2p_-\Sigma^4(\phi)}\hat{n}+p_-\hat{\tilde{n}}-i\frac{\bm{\gamma}_{\perp}\cdot\bm{x}_{\perp}}{\Sigma^2(\phi)}+m\right]u_s(\bm{0}),
\end{split}
\end{equation}
These equations indeed show that the wave packet has the same flying-focus structure of the scalar wave function in Eq. (\ref{KG_Sol_Conf}) with a transverse Gaussian profile together with a more complex spinorial prefactor.

At this point, we can consider the case of an electron in the presence of the plane wave $A^{\mu}(\phi)$, which is described by the dressed Dirac equation \cite{Landau_b_4_1982}
\begin{equation}
\label{Dirac_Eq_A}
\{\gamma^{\mu}[i\partial_{\mu}-eA_{\mu}(\phi)]-m\}\psi=0.
\end{equation}
It is known that the positive-energy solution $\psi_{V,s}(x;\bm{p})$ corresponding to an arbitrary initial momentum $\bm{p}$ and spin quantum number $s$ can be written as \cite{Volkov_1935,Landau_b_4_1982,Ritus_1985}
\begin{equation}
\psi_{V,s}(x;\bm{p})=E(x;\bm{p})u_s(\bm{p}),
\end{equation}
where the matrix
\begin{equation}
E(x;\bm{p})=e^{-i(px)-i\int_0^{\phi}d\phi'\left[\frac{e(pA(\phi'))}{p_-}-\frac{e^2A^2(\phi')}{2p_-}\right]}\left[1+\frac{e\hat{n}\hat{A}(\phi)}{2p_-}\right]
\end{equation}
is known as Ritus matrix and has, in particular, the following property
\begin{equation}
\gamma^{\mu}[i\partial_{\mu}-eA_{\mu}(\phi)]E(x;\bm{p})=E(x;\bm{p})\hat{p},
\end{equation}
where it is implicitly assumed that on the right of $E(x;\bm{p})$ on the left-hand side (and then on the right-hand side) there is no function of the coordinates.

Looking at the structure of the solution in Eq. (\ref{Dirac_Sol}), it is clear that the corresponding solution in a plane wave, is obtained by replacing the function $e^{-i(px)}$ with the Ritus matrix $E(x;\bm{p})$, i.e.,
\begin{equation}
\label{psi_FF_p}
\psi_{V,FF,s}(\phi,T,\bm{x}_{\perp})=2\pi\sigma^2\int\frac{dp_-}{2\pi}\int\frac{d^2\bm{p}_{\perp}}{(2\pi)^2}f(p_-)e^{-\bm{p}_{\perp}^2\sigma^2/2}E(\phi,T,\bm{x}_{\perp};\bm{p})u_s(\bm{p}),
\end{equation}
where, with an abuse of notation and in agreement with the left-hand side, we have written the Ritus matrix as $E(\phi,T,\bm{x}_{\perp};\bm{p})$.

Also in the present case  for $\sigma^2\gg 2/(p_-+m)^2$ the integral in $d^2\bm{p}_{\perp}$ is approximately Gaussian and can be taken analytically. It is convenient to introduce the notation
\begin{align}
M^2(\phi)&=m^2+\frac{e^2}{\phi}\int_0^{\phi}d\phi'\bm{A}_{\perp}^2(\phi'),\\
\bm{X}_{\perp}(\phi)&=\bm{x}_{\perp}+\frac{e}{p_-}\int_0^{\phi}d\phi'\bm{A}_{\perp}(\phi'),
\end{align}
such that the solution of the dressed Dirac equation for $\sigma^2\gg 2/(p_-+m)^2$ can be approximately written as
\begin{equation}
\label{Dirac_Volkov_Sol}
\begin{split}
&\psi_{V,FF,s}(\phi,T,\bm{x}_{\perp})\approx\int\frac{dp_-}{2\pi}f(p_-)e^{-i\left[\frac{M^2(\phi)}{2p_-}\phi+p_-T\right]}\frac{\sigma^2}{\Sigma^2(\phi)}e^{-\frac{\bm{X}_{\perp}^2(\phi)}{2\Sigma^2(\phi)}}\left[1+\frac{e\hat{n}\hat{A}(\phi)}{2p_-}\right]\\
&\quad\times\frac{\sqrt{p_-/m}}{p_-+m}\left[\frac{m^2\Sigma^4(\phi)+2\Sigma^2(\phi)-\bm{X}_{\perp}^2(\phi)}{2p_-\Sigma^4(\phi)}\hat{n}+p_-\hat{\tilde{n}}-i\frac{\bm{\gamma}_{\perp}\cdot\bm{X}_{\perp}(\phi)}{\Sigma^2(\phi)}+m\right]u_s(\bm{0}).
\end{split}
\end{equation}

The function $f(p_-)$ in this equation is still arbitrary provided that it vanishes for $p_-<0$. It can be chosen in order to appropriately normalize the wave function $\psi_{V,FF,s}(\phi,T,\bm{x}_{\perp})$. From the fact that the conserved four-current in the Dirac case is $\bar{\psi}(x)\gamma^{\mu}\psi(x)$ and by assuming that the function $f(p_-)$ is chosen in such a way that the wave packet vanishes for $z\to\pm\infty$, one can show that
\begin{equation}
\frac{d}{dt}\int d^3x\,\bar{\psi}(x)\gamma^0\psi(x)=0.
\end{equation}
Thus, one can impose the normalization condition for $t\to-\infty$, when the field (and the four-vector potential) is assumed to vanish. By normalizing the wave packet $\psi_{V,FF,s}(\phi,T,\bm{x}_{\perp})$ to unity, one obtains the condition
\begin{equation}
\label{Norm}
\int dp_-p_-|f(p_-)|^2=\frac{1}{\sigma^2}.
\end{equation}
It is not surprising that in the limit $\sigma\to\infty$ the function $f(p_-)$ would vanish identically. The reason is that the wave function $\psi_{V,FF,s}(\phi,T,\bm{x}_{\perp})$ would become uniform over the entire transverse plane in the same limit.

More importantly, the choice of the function $f(p_-)$ determines the pulse shape in the light-cone coordinate $T$. Analogously as in Ref. \cite{Di_Piazza_2021}, we choose the Gaussian function
\begin{equation}
\label{f}
f(p_-)=F\theta(p_-)\frac{p_-}{p_{0,-}}e^{-(p_--p_{0,-})^2\frac{\tau^2}{8}},
\end{equation}
where $F$ is a real constant to be determined from the normalization condition, $\theta(\cdot)$ is the Heaviside step function, $p_{0,-}>0$ indicates (approximately) the central value of the light-cone momentum $p_-$ and $\tau$ corresponds to the electron wave packet duration. The integral in $p_-$ in Eq. (\ref{Dirac_Volkov_Sol}) cannot be taken analytically even with such a simple pulse function. However, an approximated, explicit expression of the wave packet $\psi_{V,FF,s}(\phi,T,\bm{x}_{\perp})$ can be found under the assumption that the pulse is relatively long, i.e., that $\tau$ is large enough that the function $f(p_-)$ is well peaked around $p_{0,-}$. As a first requirement, we assume that $p_{0,-}\tau\gg 1$ such that the presence of the Heaviside function can be ignored and nevertheless the integral in $p_-$ can be taken from $-\infty$ to $+\infty$. For sufficiently large values of $\tau$ in the sense indicated above all the functions in the integral in Eq. (\ref{Dirac_Volkov_Sol}) except the phases which have to be treated with care, can be approximated with their value at $p_{0,-}$\footnote{In order to observe the FF features one typically assumes that the pulse length $\tau$ is much longer than the central Rayleigh length $l_{p_{0,-}}=p_{0,-}\sigma^2$, a more restrictive condition than the one given in the text, at least in the situations considered more in detail below, where $\sigma\lesssim \lambda_C$ and $p_{0,-}\gg m$.}. By indicating the corresponding quantities with the index 0 and by observing that the normalization condition implies (assuming $F>0$)
\begin{equation}
\label{F}
F\approx\sqrt{\frac{\tau}{2\sqrt{\pi}p_{0,-}}}\frac{1}{\sigma},
\end{equation}
we obtain
\begin{equation}
\begin{split}
&\psi_{V,FF,s}(\phi,T,\bm{x}_{\perp})\approx\sqrt{\frac{\tau}{2\sqrt{\pi}p_{0,-}}}\frac{\sigma}{\Sigma_0^2(\phi)}e^{-\frac{\bm{X}_{0,\perp}^2(\phi)}{2\Sigma_0^2(\phi)}}\left[1+\frac{e\hat{n}\hat{A}(\phi)}{2p_{0,-}}\right]\\
&\quad\times\frac{\sqrt{p_{0,-}/m}}{p_{0,-}+m}\left[\frac{m^2\Sigma_0^4(\phi)+2\Sigma_0^2(\phi)-\bm{X}_{0,\perp}^2(\phi)}{2p_{0,-}\Sigma_0^4(\phi)}\hat{n}+p_{0,-}\hat{\tilde{n}}-i\frac{\bm{\gamma}_{\perp}\cdot\bm{X}_{0,\perp}(\phi)}{\Sigma_0^2(\phi)}+m\right]u_s(\bm{0})\\
&\quad\times\int\frac{dp_-}{2\pi}e^{-(p_--p_{0,-})^2\frac{\tau^2}{8}-i\left[\frac{M^2(\phi)}{2p_-}\phi+p_-T\right]}.
\end{split}
\end{equation}
The remaining integral
\begin{equation}
I=\int\frac{dp_-}{2\pi}e^{-(p_--p_{0,-})^2\frac{\tau^2}{8}-i\left[\frac{M^2(\phi)}{2p_-}\phi+p_-T\right]}=\frac{2e^{-ip_{0,-}T}}{\tau}\int\frac{d\rho}{2\pi}e^{-\frac{\rho^2}{2}-i2\rho \frac{T}{\tau}-i\frac{M^2(\phi)\phi}{2p_{0,-}+4\rho/\tau}}
\end{equation}
can be taken by expanding the last function in the exponential for small values of $\rho$ up to terms in $\rho^2$. In fact, the variable $T$ is in general not constrained by any physical considerations and the corresponding exponential has to be taken into account exactly. On the other hand, the variable $\phi$ is naturally constrained by the Rayleigh length $l_{p_{0,-}}=p_{0,-}\sigma^2$, such that the mentioned approximation is valid if
\begin{equation}
\frac{m^2(1+\xi_0^2/2)}{p_{0,-}^2}\frac{\sigma^2}{\tau^2}\frac{1}{p_{0,-}\tau}\ll 1,
\end{equation}
where $\xi_0=|e|E_0/m\omega_0$ is the classical nonlinearity parameter of the electromagnetic plane wave  \cite{Di_Piazza_2012,Gonoskov_2022,Fedotov_2023}, with $E_0=\omega_0A_0$ being a measure of the amplitude of the plane-wave electric field and with $\omega_0$ being the central plane-wave angular frequency. In the ultrarelativistic case the above condition is easily seen not to be severely restrictive. After performing the resulting Gaussian integral, we obtain
\begin{equation}
\begin{split}
I&\approx\sqrt{\frac{2p_{0,-}^3}{\pi[p_{0,-}^3\tau^2+4iM^2(\phi)\phi]}}e^{-i\left[\frac{M^2(\phi)}{2p_{0,-}}\phi+p_{0,-}T\right]}
e^{-\frac{1}{2}\left[T-\frac{M^2(\phi)\phi}{2p_{0,-}^2}\right]^2/\left[\frac{\tau^2}{4}+i\frac{M^2(\phi)\phi}{p_{0,-}^3}\right]}\\
&\approx \sqrt{\frac{2}{\pi}}\frac{1}{\tau}e^{-i\left[\frac{M^2(\phi)}{2p_{0,-}}\phi+p_{0,-}T\right]}
e^{-\frac{2}{\tau^2}\left[T-\frac{M^2(\phi)\phi}{2p_{0,-}^2}\right]^2}
\end{split}
\end{equation}
and
\begin{equation}
\begin{split}
&\psi_{V,FF,s}(\phi,T,\bm{x}_{\perp})\approx\frac{e^{-i\left[\frac{M^2(\phi)}{2p_{0,-}}\phi+p_{0,-}T\right]}}{\sqrt{\pi^{3/2}\tau p_{0,-}}}\frac{\sigma}{\Sigma_0^2(\phi)}e^{-\frac{\bm{X}_{0,\perp}^2(\phi)}{2\Sigma_0^2(\phi)}-\frac{2}{\tau^2}\left[T-\frac{M^2(\phi)\phi}{2p_{0,-}^2}\right]^2}\left[1+\frac{e\hat{n}\hat{A}(\phi)}{2p_{0,-}}\right]\\
&\quad\times\frac{\sqrt{p_{0,-}/m}}{p_{0,-}+m}\left[\frac{m^2\Sigma_0^4(\phi)+2\Sigma_0^2(\phi)-\bm{X}_{0,\perp}^2(\phi)}{2p_{0,-}\Sigma_0^4(\phi)}\hat{n}+p_{0,-}\hat{\tilde{n}}-i\frac{\bm{\gamma}_{\perp}\cdot\bm{X}_{0,\perp}(\phi)}{\Sigma_0^2(\phi)}+m\right]u_s(\bm{0}).
\end{split}
\end{equation}
It is interesting to notice how the focus oscillates on the $x$-$y$ plane due to the presence of the plane wave (see the first Gaussian function of the transverse coordinates $\bm{x}_{\perp}$). This is a new feature as compared to the flying focus light beams, which is related here to the fact that the electron is charged and the time-evolution of the wave packet is driven by the external plane wave. Another new feature, which is also related here to the fact that the electron is charged, is that the Gaussian dependence on the variable $T$ also encodes the possibility that the electron is reflected by the plane wave if $\xi_0\gtrsim p_{0,-}/m$. Such a reflection also occurs classically according to the Lorentz equation (see also Ref. \cite{Di_Piazza_2009}, where this effect has been exploited as a possible way of detecting classical radiation reaction).

\section{Nonlinear Compton scattering by a flying focus electron}
The $S$-matrix element of nonlinear Compton scattering by a flying focus electron is conveniently written by using the expression in Eq. (\ref{psi_FF_p}) of the wave packet corresponding to a flying focus electron. At the leading order in $\alpha$, we obtain \cite{Di_Piazza_2012,Fedotov_2023}
\begin{equation}
\label{S_FF}
S_{FF}=-ie\int d^4x\,\bar{\psi}_{V,s'}(x;\bm{p}')\hat{e}^*_l(\bm{k})e^{i(kx)}\psi_{V,FF,s}(x),
\end{equation}
where $\bm{p}'$ and $s'$ ($\bm{k}$ and $l$) are the momentum and the spin (polarization) quantum number of the final electron (emitted photon). Also, $k^{\mu}=(\omega,\bm{k})$ and $e^{\mu}_l(\bm{k})$ denote the (on-shell) photon four-momentum and polarization four-vector.

By substituting Eq. (\ref{psi_FF_p}) in Eq. (\ref{S_FF}), one sees that the integrals over the light-cone coordinates $T$ and $\bm{x}_{\perp}$ results into delta functions enforcing the light-cone momentum conservation laws $p_-=p'_-+k_-$ and $\bm{p}_{\perp}=\bm{p}'_{\perp}+\bm{k}_{\perp}$. These delta functions can be exploited to take the corresponding momentum integrals in the expression of the flying-focus wave packet and the result is
\begin{equation}
\begin{split}
S_{FF}&=-2\pi ie\sigma^2e^{-\bm{p}^2_{\perp}\sigma^2/2}f(p_-)\int d\phi\,\bar{u}_{s'}(\bm{p}')\left[1-\frac{e\hat{n}\hat{A}(\phi)}{2p'_-}\right]\hat{e}^*_l(\bm{k})\left[1+\frac{e\hat{n}\hat{A}(\phi)}{2p_-}\right]u_s(\bm{p})\\
&\quad\times e^{i\int_0^{\phi}d\phi'\left\{\frac{m^2+[\bm{p}'_{\perp}-e\bm{A}_{\perp}(\phi')]^2}{2p'_-}+\frac{\bm{k}_{\perp}^2}{2k_-}-\frac{m^2+[\bm{p}_{\perp}-e\bm{A}_{\perp}(\phi')]^2}{2p_-}\right\}}.
\end{split}
\end{equation}

The corresponding differential probability on the emitted photon momentum and summed (averaged) over the final (initial) discrete quantum numbers can be computed by squaring the modulus of $S_{FF}$, by multiplying $|S_{FF}|^2$ by the phase-space factors $d^3\bm{k}/[(2\pi)^32\omega]$ and $d^3\bm{p}'/[(2\pi)^32\varepsilon_{p'}]$, and by using the relations $\sum_s u_s(\bm{p})\bar{u}_s(\bm{p})=\hat{p}+m$ and $\sum_le^{\mu}_l(\bm{k})e^{*,\nu}_l(\bm{k})\to -\eta^{\mu\nu}$ \cite{Landau_b_4_1982}, the latter substitution rule exploiting the gauge invariance of QED. By proceeding analogously as in, e.g., Ref. \cite{Di_Piazza_2018}, the result is
\begin{equation}
\begin{split}
dP_{FF}&=-\alpha 4m^2\sigma^4\frac{d^3\bm{k}}{(2\pi)^3}\frac{1}{2\omega}\int\frac{d^3\bm{p}'}{2\varepsilon_{p'}}|f(p_-)|^2e^{-\bm{p}^2_{\perp}\sigma^2}\int d\phi d\phi'\,\mathcal{T}\\
&\quad\times e^{i\int_{\phi'}^{\phi}d\tilde{\phi}\left\{\frac{m^2+[\bm{p}'_{\perp}-e\bm{A}_{\perp}(\tilde{\phi})]^2}{2p'_-}+\frac{\bm{k}_{\perp}^2}{2k_-}-\frac{m^2+[\bm{p}_{\perp}-e\bm{A}_{\perp}(\tilde{\phi})]^2}{2p_-}\right\}},
\end{split}
\end{equation}
where (see Ref. \cite{Di_Piazza_2018})
\begin{equation}
\begin{split}
\mathcal{T}&=-\frac{1}{4}\text{Tr}\left\{(\hat{p}'+m)\bigg[1-\frac{e}{2 p'_-}\hat{n}\hat{A}(\phi)\bigg]
\gamma^{\mu}\bigg[1+\frac{e}{2 p_-}\hat{n}\hat{A}(\phi)\bigg](\hat{p}+m)\right.\\
&\quad\left.\times\bigg[1-\frac{e}{2 p_-}\hat{n}\hat{A}(\phi')\bigg]\gamma_{\mu}\bigg[1+\frac{e}{2 p'_-}\hat{n}\hat{A}(\phi')\bigg]\right\},
\end{split}
\end{equation}
and where the quantities $p_-$ and $\bm{p}_{\perp}$ are understood to be equal to $p'_-+k_-$ and $\bm{p}'_{\perp}+\bm{k}_{\perp}$, respectively. 

Since the trace $\mathcal{T}$ is identical to the one in Ref. \cite{Di_Piazza_2018} of nonlinear Compton scattering by an electron with momentum $\bm{p}$, we can use Eqs. (7) and (8) of Ref. \cite{Di_Piazza_2018} to write the differential probability as
\begin{equation}
\label{dP_d^3k}
\begin{split}
dP_{FF}&=-\frac{d^3\bm{k}}{(2\pi)^3}\frac{4\alpha m^2\sigma^4}{2\omega}\int\frac{d^3\bm{p}'}{\varepsilon_{p'}}|f(p_-)|^2e^{-\bm{p}^2_{\perp}\sigma^2}\\
&\quad\times\int d\phi d\phi'\,e^{-i\frac{k_-m^2}{2p_-p'_-}\int_{\phi}^{\phi'}d\tilde{\phi}\,\left\{1+\left[\frac{\bm{p}_{\perp}}{m}-\frac{p_-}{k_-}\frac{\bm{k}_{\perp}}{m}-\bm{\xi}_{\perp}(\tilde{\phi})\right]^2\right\}}\left\{1+\frac{1}{4}\frac{p_-^2+p_-^{\prime\,2}}{p_-p'_-}[\pmb{\xi}_{\perp}(\phi)-\pmb{\xi}_{\perp}(\phi')]^2\right\},
\end{split}
\end{equation}
where $\pmb{\xi}_{\perp}(\phi)=e\bm{A}_{\perp}(\phi)/m$. Note that in the limit $\sigma\to\infty$ this expression of the emission probability tends to the corresponding expression for an incoming electron wave packet described by the function $f(p_-)$ with the correct normalization given in Eq. (\ref{Norm}). In agreement with the results in Ref. \cite{Angioi_2016}, we see that the emission probability in Eq. (\ref{dP_d^3k}) is given by the average over the momenta distribution of the wave packet of the corresponding probability for an electron with a definite momentum. This has also an interesting consequence, which is related to the fact that we are considering here a plane wave as background field: The probability $dP_{FF}$ does not depend on the value of $\phi$ at which the flying-focus wave packet is focused. Indeed, changing the position of the focus corresponds to shift the variable $\phi$ in the function $\Sigma(\phi)$ (see Eq. (\ref{Sigma_sq})) by a constant $\phi_0$. This in turn alters the expression of the flying-focus wave packet by a constant (in configuration space) phase $\exp(-ip_+\phi_0)$, which does not affect the momenta distribution.

Equation (\ref{dP_d^3k}) is our main result although its numerical evaluation is nontrivial due to the additional three-dimensional integral over the initial electron wave packet as compared to the case of an incoming electron with definite momentum (see Ref. \cite{Angioi_2016} for a study of the collision of an electron wave packet, described by a three-dimensional Gaussian in momentum space, with an intense laser beam). Notice that by using the identity $\int d^3\bm{p}'/\varepsilon_{p'}=\int dp'_-d^2\bm{p}'_{\perp}/p'_-$, and changing variables to $p_-=p'_-+k_-$ and $\bm{p}_{\perp}=\bm{p}'_{\perp}+\bm{k}_{\perp}$, the integral over $d^2\bm{p}_{\perp}$ turns out to be Gaussian and can be taken analytically. The remaining extra integral in $dp_-$ can be taken approximately in the case of long electron pulses, as described above. However, the remaining integrals in $\phi$ and $\phi'$ would still contain divergences that need to be eliminated by standard methods (see, e.g. Ref. \cite{Mackenroth_2011}) although the regularization of these integrals remains non-trivial. 

Now, by isolating the quadratic terms in $\bm{p}_{\perp}$ in the exponent in Eq. (\ref{dP_d^3k}) and looking at the coefficient of $\bm{p}_{\perp}^2$ already gives us the possibility of identifying the important parameter that controls the effects of the electron focusing on the emission probability (a more quantitative condition on the importance of these effects is obtained below, within the so-called locally-constant field approximation (LCFA)). One sees that this is the case, i.e., that electron wave-packet focusing effects are important, for
\begin{equation}
\sigma\lesssim\lambda_C\sqrt{\frac{k_-}{p'_-}\frac{m^2\phi_-}{2p_-}},
\end{equation}
where $\phi_-=\phi-\phi'$. The values of the coordinate $\phi_-$ are effectively limited by the formation length of the process and $|\phi_-|\lesssim m/|e|E_0$ \cite{Ritus_1985,Baier_b_1998,Di_Piazza_2018}. This allows one to conclude that
\begin{equation}
\sigma\lesssim\lambda_C\sqrt{\frac{k_-}{2\chi_0p'_-}},
\end{equation}
where $\chi_0=(p_-/m)(E_0/E_{cr})$ is the quantum nonlinearity parameter (corresponding to the momentum component $p_-$ of the electron wave packet) and $E_{cr}=m^2/|e|$ is the critical field of QED \cite{Di_Piazza_2012,Gonoskov_2022,Fedotov_2023}. In this way, we see that for any value of $\chi_0$ the effects of the focal properties of the electron wave packet on the non-negligible region of the emission spectrum ($k_-\lesssim \chi_0p_-$ for $\chi_0\ll 1$ and $k_-\lesssim p_-$ for $\chi_0\gtrsim 1$) are important for values of $\sigma$ of the order of the Compton wavelength.

For the above reason, we content ourselves here with the numerical evaluation of the differential probability within the LCFA, which is generally speaking valid for $\xi_0\gg 1$ (see Refs. \cite{Di_Piazza_2018,Di_Piazza_2019,Alexandrov_2019,Ilderton_2019_b,Podszus_2019,Ilderton_2019} for a thorough discussion about the validity of this approximation). Within this approximation, it is convenient to introduce explicitly the typical angular frequency $\omega_0$ characterizing the background plane wave and to express the double integral in $d\varphi d\varphi'=\omega_0^2d\phi d\phi'$ as a double integral in the average variable $\varphi_+=(\varphi+\varphi')/2$ and the relative variable $\varphi_-=\varphi-\varphi'$ because the integrand can be then expanded for small values of $|\varphi_-|$. Due to the similarity of Eq. (\ref{dP_d^3k}) with the corresponding result for an initial electron with definite momentum $\bm{p}$, we can use Eqs. (40)-(43) in Ref. \cite{Di_Piazza_2019} to conclude that the differential probability the differential probability $dP_{FF,\text{LCFA}}/d\bm{k}$ can be written as $dP_{FF,\text{LCFA}}/d\bm{k}=\int d\varphi_+\,dP_{FF,\text{LCFA}}/d\bm{k}d\varphi_+$, where
\begin{equation}
\label{dP_d^3k_dphi_LCFA}
\begin{split}
\frac{dP_{FF,\text{LCFA}}}{d\bm{k}d\varphi_+}&=\frac{\alpha\sigma^4}{\sqrt{3}\pi^3\omega_0\omega}\int \frac{dp'_-d^2\bm{p}'_{\perp}}{p'_-}p_-|f(p_-)|^2e^{-\bm{p}^2_{\perp}\sigma^2}\frac{\sqrt{1+\bm{\pi}^2_{\perp}(\varphi_+)}}{\chi(\varphi_+)}\\
&\quad\times\left\{\frac{p_-^2+p_-^{\prime\,2}}{p_-p'_-}\left[1+\bm{\pi}^2_{\perp}(\varphi_+)\right]-1\right\}\text{K}_{1/3}\left(\frac{2}{3}\frac{k_-}{p'_-}\frac{\left[1+\bm{\pi}^2_{\perp}(\varphi_+)\right]^{3/2}}{\chi(\varphi_+)}\right),
\end{split}
\end{equation}
with $\chi(\varphi_+)=(\omega_0 p_-/m^2)|\bm{\xi}'_{\perp}(\varphi_+)|$,
\begin{equation}
\bm{\pi}_{\perp}(\varphi_+)=\frac{\bm{p}_{\perp}}{m}-\frac{p_-}{k_-}\frac{\bm{k}_{\perp}}{m}-\bm{\xi}_{\perp}(\varphi_+),
\end{equation}
and $\text{K}_{\nu}(\cdot)$ being the modified Bessel function of order $\nu$ \cite{NIST_b_2010}.

A comparison with the corresponding differential probability $dP_{\text{LCFA}}(\bm{p})/d\bm{k}d\varphi_+$ per unit phase for an electron with a fixed incoming momentum $\bm{p}$ (see Eqs. (40)-(43) in Ref. \cite{Di_Piazza_2019}) also shows that 
\begin{equation}
\frac{dP_{FF,\text{LCFA}}}{d\bm{k}d\varphi_+}=\frac{\sigma^4}{\pi}\int dp_-d^2\bm{p}_{\perp}p_-|f(p_-)|^2e^{-\bm{p}^2_{\perp}\sigma^2}\frac{dP_{\text{LCFA}}(\bm{p})}{d\bm{k}d\varphi_+},
\end{equation}
where the changes of variables $p_-=p'_-+k_-$ and $\bm{p}_{\perp}=\bm{p}'_{\perp}+\bm{k}_{\perp}$ have been made.

The normalization condition in Eq. (\ref{Norm}) shows that the photon yield is comparable with the case of an incoming electron with definite momentum. Moreover, in the limit $\sigma\to\infty$, one obtains
\begin{equation}
\frac{dP_{FF,\text{LCFA}}}{d\bm{k}d\varphi_+}=\sigma^2\int \frac{dp_-}{2\pi}p_-|f(p_-)|^2\frac{dP_{\text{LCFA}}(\bm{p})}{d\bm{k}d\varphi_+},
\end{equation}
which, again, shows that the resulting probability is the wave-packet average of the corresponding probability for an electron with definite (longitudinal) momentum.

In order to deduce for which values of $\sigma$ the emission probability is significantly different from the plane-wave electron case, one can compare the argument of the Bessel function and the Gaussian controlling the transverse momentum distribution of the flying focus electron. One then obtains the condition that electron flying focus effects are important for $k_-/\chi_0p'_-\gtrsim (m\sigma)^3$, i.e., for
\begin{equation}
\label{sigma}
\sigma\lesssim\lambda_C\left(\frac{k_-}{\chi_0p'_-}\right)^{1/3}.
\end{equation}
Notice that, as we have already mentioned, considering situations where $\chi_0\ll 1$ would not help to increase the upper limit of the relevant values of $\sigma$ because in that (classical) situation, the typical values of $k_-$, where the probability is significantly different from zero are $k_-\lesssim\chi_0p_-$. The result in Eq. (\ref{sigma}) can also be understood qualitatively by observing that one expects a significant effect of the focusing when the transverse momenta within the wave packet are comparable with the transverse momenta acquired by the electron in the plane wave within the formation phase of radiation. By using the expression of the formation phase $\delta\varphi_f$ as found in Ref. \cite{Di_Piazza_2018}, i.e., $\delta\varphi_f\sim (p'_-\chi_0/k_-)^{1/3}/\xi_0$, one sees that the transverse momentum acquired by the electron within $\delta\varphi_f$ is exactly $m(p'_-\chi_0/k_-)^{1/3}$, in agreement with Eq. (\ref{sigma}).

Although in the present case we only require localizing the electron on the transverse plane, it has been debated whether in principle it is possible to localize an electron down to a Compton wavelength (see Ref. \cite{Sebens_2020} for a recent publication on the topic, containing also a clear historical overview). The reason of the debate is that in order to localize the electron at such short lengths one would need, according to Heisenberg uncertainty principle, energies larger than the electron rest energy, allowing in this way for pair production \cite{Bjorken_b_1964,Landau_b_4_1982,Heitler_b_1984}. This idea has been formulated mathematically by proving that in order to localize an electron wave packet down to a Compton wavelength one necessarily needs also negative-energy states \cite{Bjorken_b_1964}. However, it has been first shown in Ref. \cite{Bracken_1999} that it is possible to construct electron wave packets only containing positive-energy states and yet localized below the Compton wavelength. Although such states do not lead to pair production, they are found to delocalize very rapidly in time \cite{Bracken_2005}. Finally, it has also been shown that the construction of states localized below the Compton scale does not lead to any contradiction within quantum field theory \cite{Pavsic_2018} (see also Ref. \cite{Angioi_2016}).

The dependence of $dP_{FF,\text{LCFA}}/d\bm{k}d\varphi_+$ on $\bm{k}_{\perp}$ can be significantly simplified by making the changes of variables $p_-=p'_-+k_-$ and $\bm{p}_{\perp}=\bm{p}'_{\perp}+\bm{k}_{\perp}$:
\begin{equation}
\begin{split}
\frac{dP_{FF,\text{LCFA}}}{d\bm{k}d\varphi_+}&=\frac{\alpha \sigma^4}{\sqrt{3}\pi^3\omega_0\omega}\int_{k_-}^{\infty} dp_-\int d^2\bm{p}_{\perp}\frac{p_-|f(p_-)|^2}{p_--k_-}e^{-\bm{p}^2_{\perp}\sigma^2}\frac{\sqrt{1+\bm{\pi}^2_{\perp}(\varphi_+)}}{\chi(\varphi_+)}\\
&\quad\times\left\{\frac{p_-^2+(p_--k_-)^2}{p_-(p_--k_-)}\left[1+\bm{\pi}^2_{\perp}(\varphi_+)\right]-1\right\}\text{K}_{1/3}\left(\frac{2}{3}\frac{k_-}{p_--k_-}\frac{\left[1+\bm{\pi}^2_{\perp}(\varphi_+)\right]^{3/2}}{\chi(\varphi_+)}\right).
\end{split}
\end{equation}
Finally, by choosing the function $f(p_-)$ as in Eq. (\ref{f}) and by assuming that we look at photon light-cone momenta $k_-$ such that $p_{0,-}-k_-\gg 1/\tau$, we have
\begin{equation}
\label{dP_d^3k_dphi_LCFA_F}
\begin{split}
\frac{dP_{FF,\text{LCFA}}}{d\bm{k}d\varphi_+}&=\frac{\alpha\sigma^2}{\sqrt{3}\pi^3\omega_0\omega(p_{0,-}-k_-)}\int d^2\bm{p}_{\perp}e^{-\bm{p}^2_{\perp}\sigma^2}\frac{\sqrt{1+\bm{\pi}^2_{0,\perp}(\varphi_+)}}{\chi_0(\varphi_+)}\\
&\quad\times\left\{\frac{p_{0,-}^2+(p_{0,-}-k_-)^2}{p_{0,-}(p_{0,-}-k_-)}\left[1+\bm{\pi}^2_{0,\perp}(\varphi_+)\right]-1\right\}\\
&\quad\times\text{K}_{1/3}\left(\frac{2}{3}\frac{k_-}{p_{0,-}-k_-}\frac{\left[1+\bm{\pi}^2_{0,\perp}(\varphi_+)\right]^{3/2}}{\chi_0(\varphi_+)}\right),
\end{split}
\end{equation}
where, again, all the quantities with the index $0$ indicate that they are computed by setting $p_-=p_{0,-}$. The above condition is clearly not limiting in the case of sufficiently long electron pulses. Also, notice that Eq. (\ref{dP_d^3k_dphi_LCFA_F}) tends to the known result in Ref. \cite{Di_Piazza_2019} in the limit $\sigma\to\infty$.

\section{Numerical results and discussion}

The possibility of manipulating single-electron wave packets has attracted a lot of attention not only theoretically \cite{Reinhardt_2020} but also experimentally \cite{Ehberger_2018,Eldar_2022,Madan_2022,Tsarev_2023} (see also the reviews \cite{Baum_2013,Morimoto_2022} and the references therein). In particular, in Ref. \cite{Madan_2022} a technique is investigated to manipulate experimentally the transverse structure of an electron wave packet. Although the manipulation still occurs on length scales of the order of micrometers, one can foresee that control over smaller scales could be reached in the future. 

Below, we provide results of a numerical simulation of the collision of a flying focus electron with a plane wave. We consider an electron with initial central energy of $5\;\text{GeV}$ and with a transverse width of $\sigma=\lambda_C/2$. The plane wave propagates along the positive $z$ direction and it is assumed to be linearly polarized along the $x$ direction. The temporal profile is a Gaussian centered at $\phi=0$, with full-width half maximum of the intensity $\tau_L=20\;\text{fs}$ and peak intensity $I_0=10^{20}\;\text{W/cm$^2$}$. With these parameters it is $\xi_0\approx 4.8$ and $\chi_0\approx 0.29$, such that we expects that both nonlinear and quantum effects are important. Moreover, the LCFA is expected to be sufficiently accurate for the purposes below.

\begin{figure}
\begin{center}
\includegraphics[width=0.9\textwidth]{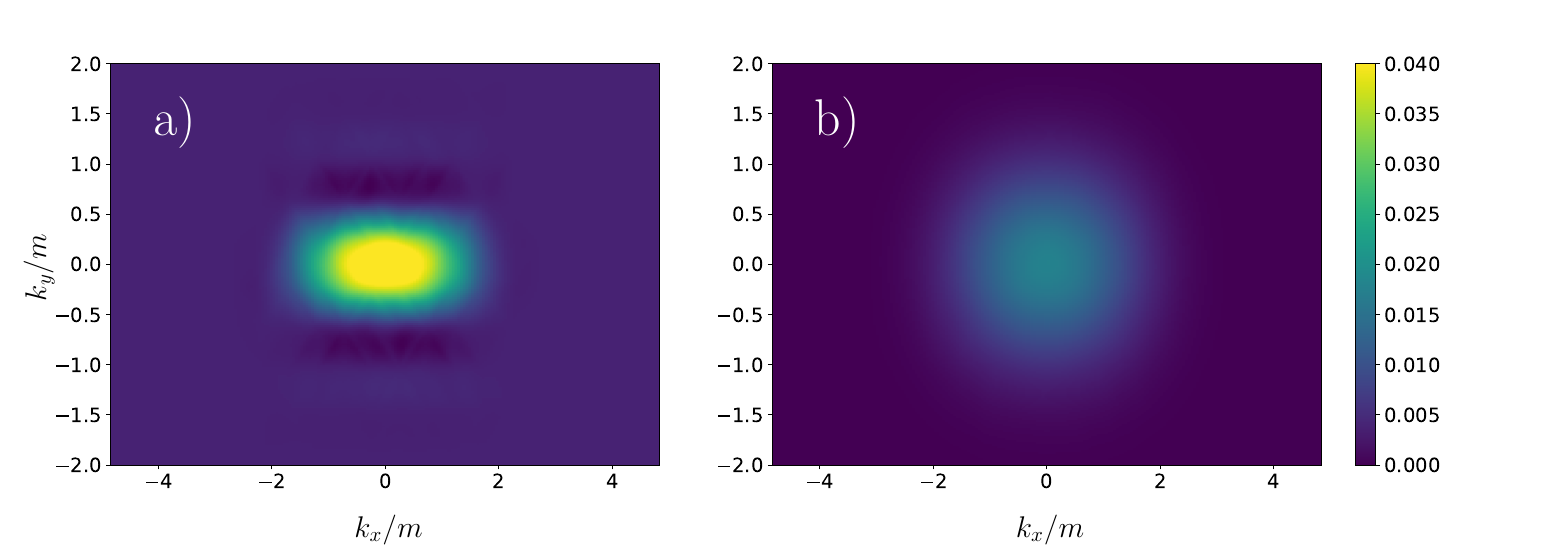}
\end{center}
\caption{Probability density of nonlinear Compton scattering by a plane-wave (part a)) and a flying-focus (part b)) electron head-on colliding with an intense plane wave. The numerical parameters are given in the text.}
\label{Figure_1}
\end{figure}

In Fig. \ref{Figure_1} the photon emission probability is shown as a function of the transverse components $k_x/m$ (horizontal axis) and $k_y/m$ (vertical axis) of the emitted photon for $k_-=0.452\,p_-\approx 4.5\;\text{GeV}$. 
The figure shows that the main difference between the two spectra is a wider emission along the magnetic field direction, i.e., the $y$ direction. This is due to the focusing of the flying-focus electron whose wave packet includes transverse momenta typically exceeding the electron mass. On the other hand, along the polarization direction of the plane wave, the radiation extends typically to momenta of the order of $m\xi_0$ and therefore there is not much difference between the two situations (notice that the scales of the horizontal and vertical axes in Fig. \ref{Figure_1} are different). As a result, the maximum value of the probability density is somewhat lower in the case of the flying-focus electron than in the case of a plane-wave electron. We have verified numerically that by considering tighter flying-focus beams the difference in the emission width along the magnetic field of the plane wave increases. However, no additional qualitative differences are observed. The above discussion hints to the fact that, by considering a plane wave with $\xi_0\sim 1$, one would expect effects of the focusing of the electron along both transverse directions, which would however require much higher electron energies to increase the total signal and especially if one wants to work within the quantum regime $\chi_0\sim 1$. 

\section{Conclusions and outlook}

In this paper we have considered the emission of radiation by a flying-focus electron wave packet in an intense plane-wave field. In order to compute the spectrum, we have first found a new solution of the Dirac equation in the presence of an arbitrary plane wave, representing an electron wave packet, whose focal region moves in the opposite direction as the electron average velocity (assuming that the electron is not reflected by the electromagnetic plane wave). We have determined analytically the radiation emission spectrum in general and in closed form within the locally-constant crossed field approximation. Results from numerical simulations in the case of a linearly-polarized plane wave show that for electron wave packets focused below a Compton wavelength the radiation width along the magnetic field of the plane wave turns out to be larger than in the case of a plane-wave electron exactly counterpropagating with respect to the plane wave. The reason for the asymmetry on the transverse plane is that the electron emits photons with momenta up to $m\xi_0$ along the plane-wave polarization and $\xi_0$ is assumed to be much larger than unity within the locally-constant field approximation. Thus, by considering a plane wave with $\xi_0\sim 1$ one would expect effects of the electron focusing along both transverse directions. This, would however require much higher electron energies in order to stay within the regime where $\chi\sim 1$. The discussed effects on the angular distribution of the emitted photons could represent a signature that a flying-focus wave function has been realized experimentally.

The discussed scenario is amenable of an essentially complete analytical treatment due to its relative simplicity but it may not be the ideal one to observe the effects of the flying-focus nature of the electron wave packet. Another possibility to observe such effects can be to introduce a new length scale in the problem, e.g., to consider a focused laser field. We note, however, that at least in the regime where the initial average energy of the electron is much larger than $m\xi_0$, the transverse formation length of nonlinear Compton scattering is typically of the order of the Compton wavelength \cite{Di_Piazza_2021}. Thus, we expect that including the effects of the laser focusing will amount to additionally average the results over the laser pulse shape. In this respect, it can be interesting to study the interaction of a flying-focus electron wave packet with a flying-focus laser beam, as in this case the effects of the transverse formation length are expected to be enhanced, according to the results in Ref. \cite{Di_Piazza_2021}. Another appealing configuration could be to consider electron average energies comparable with $m\xi_0$ and only mildly relativistic ($\xi_0\sim 1$) such that the electron is expected to span a region on the transverse plane of the order of the laser focal region, if the latter is tightly focused. These are all possible promising scenarios, which, however, require further analysis for providing quantitative answers and are left for future investigation.

\begin{acknowledgments}
This material is based upon work supported by the U.S. Department of Energy [National Nuclear Security Administration] University of Rochester ``National Inertial Confinement Fusion Program'' under Award Number DE-NA0004144 and U.S. Department of Energy, Office of Science, under Award Number DE-SC0021057.

This report was prepared as an account of work sponsored by an agency of the United States Government. Neither the United States Government nor any agency thereof, nor any of their employees, makes any warranty, express or implied, or assumes any legal liability or responsibility for the accuracy, completeness, or usefulness of any information, apparatus, product, or process disclosed, or represents that its use would not infringe privately owned rights. Reference herein to any specific commercial product, process, or service by trade name, trademark, manufacturer, or otherwise does not necessarily constitute or imply its endorsement, recommendation, or favoring by the United States Government or any agency thereof. The views and opinions of authors expressed herein do not necessarily state or reflect those of the United States Government or any agency thereof.

The work of M.F. is supported by the European Union’s Horizon Europe research and
innovation program under the Marie Sklodowska-Curie Grant Agreement No. 101105246-STEFF.

\end{acknowledgments}

%


\end{document}